\documentclass[12pt]{revtex4}
\usepackage{graphicx}
\begin{document}

\title{Productions of Heavy Flavored Mesons in Relativistic Heavy Ion Collisions in the Recombination Model}
\author{R. Peng$^{a,b,c}$ and C. B. Yang$^{a,b}$}
\affiliation{$^{a}$Institute of Particle Physics, Hua-Zhong Normal
University, Wuhan 430079, People's Republic of China\\
$^{b}$ Key Laboratory of Quark and Lepton (Hua-Zhong Normal University), Ministry of Education,
Wuhan 430079, People's Republic of China\\
$^c$College of Science, Wuhan University of Science and Technology,Wuhan 430065, People's Republic of China\\
$^a$ Corresponding author, E-mail address: pengru$\_$1204@hotmail.com.}

\begin{abstract}
We get the distributions of shower partons initiated by heavy quarks $c$ and $b$ by studying the fragmentation functions in the framework of the recombination model. The transverse momentum spectra of heavy flavored mesons are predicted with these distributions. We find that the contribution from the recombination of thermal-shower partons is an important part in the total spectrum for the mesons. We predict the heavy flavored meson productions for different centralities with the heavy quark fugacities fitted by the experimental data of $J/\psi$ transverse momentum spectra in Au+Au collisions.



\end{abstract}
\maketitle

\section{Introduction}
Heavy quark production in relativistic heavy ion collisions is a subject of current interest for understanding the quark-gluon interactions. Since heavy quarks are produced in the initial hard scattering processes at RHIC, they are sensitive to probe the medium formed in the collisions as they may lose energy by gluon radiation during propagating through the medium. Much progress has been made to improve theoretical understanding of the production mechanism of heavy mesons containing a heavy quark ($c$ or $b$ quark), such as the discovery of the powerful heavy quark spin-flavor symmetry \cite{heavy} and the development of the Heavy Quark Effective Theory (HQET) \cite{HQET} based on QCD. But the techniques of HQET can not be applied to $B_c(\overline{b}c)$ meson since the charm quark is not sufficiently light. There are two mechanisms investigated for the production of $b$-flavored mesons: recombination \cite{recombination} and fragmentation. And it has been pointed out \cite{fragmentation} that the dominant mechanism is due to fragmentation, especially in the region of large transverse momentum. In Ref.\cite{pengru1}, we have shown the important contribution from the recombination of thermal-shower partons to the transverse momentum spectrum of $J/\psi$ in central Au+Au collisions.

Up to now, the recombination model for particle production has achieved a great success. Especially the fragmentation process has been explained as a recombination of semi-hard shower partons in a jet. The model displays how the final state particles are produced from the recombination of thermal and shower partons created by hard partons. The determination of the shower parton distributions (SPD) has been obtained by studying the fragmentation functions (FF) in the framework of the recombination model \cite{pengru1,SPD}. With the parametrized  SPDs, the model has offered good agreement with the experimental data on the hadron spectra from low to intermediate $p_T$ \cite{hadron}, and successfully explained the Cronin effect on pion production in d+Au collisions \cite{Cronin}. Also, based on the good description of light hadrons productions, the model has been extended to the productions of strange hadrons \cite{strange} and $J/\psi$ \cite{pengru1} in central Au+Au collisions. In this paper we complete the SPDs initiated by hard partons of $c$ and $b$ quarks and furthermore predict the transverse momentum spectra for heavy-flavored mesons.

This work is organized as follows. We first review the basic formulism for the meson production in the recombination model in the next section. Then we show the results of charmed mesons and $b$-flavored mesons productions in central Au+Au collisions at $\sqrt{s_{NN}}=200$ GeV in Section 3 and 4, respectively. In Section 5 we predict the heavy flavored meson transverse momentum spectra for non-central centralities. A brief summary is given in the final section.

\section{Meson production in the recombination model}
The contribution to the transverse momentum spectra of meson consists three terms because there are two components of parton sources: thermal ($\mathcal{T}$) partons and shower ($\mathcal{S}$) partons which are originated from hard partons. In this paper we substitute $p$ for $p_T$ to denote the transverse momentum of the meson. Then the production can be expressed as the sum of three terms
\begin{equation}
\label{production}
\frac{dN_{M}}{d^2p}=\frac{d(N_{M}^{\mathcal{TT}}+N_{M}^{\mathcal{TS}}
+N_{M}^{\mathcal{SS}})}{d^2p}.
\end{equation}
Ref.\cite{RBC} provides a theoretical description of the recombination process in a dense parton phase, where the pure thermal contribution ($\mathcal{TT}$) to meson spectrum can be written as
\begin{equation}
\label{TT}
\frac{dN^{\mathcal{TT}}_{M}}{d^{2}p}=C_{M}M_{T}\frac{\tau A_{T}}{(2\pi)^3}2\gamma_{a}\gamma_{b}I_{0}[\frac{p\mathrm{sinh}\eta_T}{T}]
\int^1_0dx\arrowvert\phi_{M}(x)\arrowvert^2k_{M}(x,p).
\end{equation}
Here
\begin{equation}
k_{M}(x,p)=K_{1}[\frac{\mathrm{cosh}\eta_T}{T}(\sqrt{m^{2}_{a}+p_{1}^2}+
\sqrt{m^{2}_{b}+p_{2}^2})],
\end{equation}
with the momenta of the two constituent quarks $p_1=xp$ and $p_2=(1-x)p$. $I_0$ and $K_1$ are the modified Bessel functions. $M_T$ is the transverse mass of the meson and $A_{T}=\rho^{2}_0\pi$ is the transverse area of the parton system with the radius $\rho_0=9$ fm \cite{RBC}. It was assumed that hadronization occurs at $\tau=5$ fm with temperature $T=175$ MeV in the parton phase \cite{RBC}, which is consistent with predictions of the phase transition temperature at vanishing baryon chemical potential from lattice QCD \cite{QCD}. The transverse flow rapidity
$\eta_T$ is defined by a flow velocity with $v_T=\mathrm{tanh}\eta_T$. $\gamma_a$ and $\gamma_{b}$ stand for the fugacities of the constituent quarks $a$ and $b$, respectively. In Ref.\cite{RBC} the fugacities of light quarks are $\gamma_u=\gamma_d=1$ and for strange quarks $\gamma_s=0.8$. Every quark has 3 color and 2 spin degrees of freedom. So we use the meson degeneracy factor $C_{M}=(3\times 2)^2$.

$\phi_{M}(x)$ in Eq.(\ref{TT}) is the wave function of the meson in the momentum space. We can describe the wave function from the definition of the recombination function (RF). The RF for the process of $q\overline{q'}\rightarrow M$ with quark $q$ at momentum fraction $x_1$ and $\overline{q'}$ at $x_2$ to form a meson at $x$ is expressed as \cite{RCH1}
\begin{equation}
\label{MRF}
R_{M}(x_{1},x_{2},x)=\frac{1}{B(a,b)}\left(\frac{x_1}{x}\right)^a\left(\frac{x_2}{x}\right)^b
\delta(\frac{x_1}{x}+\frac{x_2}{x}-1),
\end{equation}
where $B(a,b)$ is the beta function. The RF $R_M$ is an invariant distribution which is related to the non-invariant probability $G_{M}(y_1,y_2)$ of finding the two constituent quarks in $M$ with momentum fractions $y_1$ and $y_2$ through \cite{RCH1}
\begin{equation}
R_{M}(x_1,x_2,x)=y_1y_2G_M(y_1,y_2), y_i=x_i/x.
\end{equation}
The normalization factor $1/B(a,b)$ in Eq.(\ref{MRF}) is determined by the following requirement
\begin{equation}
\int^{1}_{0}dy_1\int^{1-y_1}_{0}dy_2G_M(y_1,y_2)=1.
\end{equation}
In fact, $G_M(y_1,y_2)$ relates with $\arrowvert\phi_{M}(y_1)\arrowvert^2$ by $G_M(y_1,y_2)=\arrowvert\phi_{M}(y_1)\arrowvert^2\delta(y_1+y_2-1)$. Consequently, the wave function of a meson in the momentum space corresponding to $G_{M}(y_1,y_2)$ can be written as
\begin{equation}
\label{wavefunction}
\arrowvert\phi_{M}(y_1)\arrowvert^2=\frac{1}{B(a,b)}y^{a-1}_{1}(1-y_{1})^{b-1}.
\end{equation}
For a meson, the ratio of the average momentum fraction is proportional to the masses of the constituent quarks $a/b=\overline{x_1}/\overline{x_2}=m_{a}/m_{b}$\cite{RCH2}, i.e., for $D^0$ meson, $m_c\simeq1.5$ GeV, $m_{u,d}\simeq0.3$ GeV, $a/b=m_{u}/m_{c}\simeq1/5$.

The first step of calculating the thermal-shower recombination term ($\mathcal{TS}$ contribution) is to get the shower parton distribution. Using the parametrized distribution $S^{j}_{i}(z)$ \cite{pengru1,SPD}, we can determine the distribution of shower parton $j$ with transverse momentum $p$ in central Au+Au collisions as \cite{energyloss}
\begin{equation}
\label{shower}
\mathcal{S}(p)=\sum_{i}\int\frac{dq}{q}F_{i}(q)S^{j}_{i}(p/q),
\end{equation}
where
\begin{equation}
F_{i}(q)=\frac{1}{\beta L}\int^{qe^{\beta L}}_{q}\frac{dk}{k}f'_{i}(k),
\end{equation}
with $f'_{i}(k)=f_{i}(k)\cdot(2\pi)^3/E$ \cite{pengru1}. The distribution $f_{i}(k)=dN^{hard}_{i}/d^2kdy$ of hard parton $i$ just after hard scattering in Au+Au collisions at $\sqrt{s_{NN}}=200$ GeV at mid-rapidity can be found in Refs.\cite{FGLUON} and \cite{FCHARM}. $\beta L$ is the explicit dynamical medium factor to describe the energy loss effect with $\beta L=2.39$ for $i=$light quarks, gluon in Au+Au collisions for $0-20\%$ centrality \cite{energyloss}. Recent measurements of the transverse momentum distributions and nuclear modification factors of non-photonic electrons from heavy quark decays at high $p_T$ show a suppression level of heavy quarks similar to light quarks \cite{sameL}. So we choose the same energy loss factor $\beta L$ for $i=c, b$ as that for light quarks.

In Ref.\cite{RBC}, the $\mathcal{TS}$ term can be calculated as
\begin{eqnarray}
\label{TS1}
\frac{dN^{\mathcal{TS}}_{M}}{d^{2}p}&=&C_{M}\int_{\Sigma}d\sigma_{R}
\frac{p^{\mu} u_{\mu}(R)}{(2\pi)^3}\int^{1}_{0}dx\arrowvert\phi_{M}(x)\arrowvert^2\nonumber\\
&  &[\omega_{a}(R,p_1)\mathcal{S}_{b}(R,p_2)+ \mathcal{S}_{a}(R,p_1)\omega_{b}(R,p_2)],
\end{eqnarray}
with the phase space distribution $\omega(R,p)$ of thermal partons. $u_{\mu}(R)$ is the future oriented unit vector orthogonal to the hypersurface $\Sigma$ defined by the hadronization volume. Because the thermal parton distribution density with momentum $p$ is defined as \cite{RBC}
\begin{equation}
\mathcal{T}(p)=\frac{g\gamma}{\tau A_T}\int_{\Sigma}d\sigma_{R}\frac{p^{\mu}u_{\mu}(R)}{(2\pi)^3}\omega(R,p),
\end{equation}
with $g=6$ coming from the color and spin degeneracy of a quark, we can rewrite Eq.(\ref{TS1}) as
\begin{equation}
\label{TS2}
\frac{dN^{\mathcal{TS}}_{M}}{d^{2}p}=C_{M}\int^{1}_{0}dx\arrowvert\phi_{M}(x)\arrowvert^2
[\frac{\mathcal{T}_{a}(p_1)\mathcal{S}_{b}(p_2)}{g\gamma_{a}x}
+\frac{\mathcal{S}_{a}(p_1)\mathcal{T}_{b}(p_2)}
{g\gamma_{b}(1-x)}].
\end{equation}
We get the thermal parton transverse momentum distribution density from the thermal parton spectrum given in Ref.\cite{RBC} by
\begin{equation}
\mathcal{T}(p)=\frac{dN^{th}}{dp^{2}dy}\bigg|_{y=0}\bigg/(\tau A_T ).
\end{equation}

The shower-shower recombination term ($\mathcal{SS}$ contribution) is equivalent to the FF \cite{energyloss}
\begin{equation}
\label{SS}
\frac{dN^{\mathcal{SS}}_{M}}{pdp}=\frac{1}{p^{0}p}\sum_{i}\int\frac{dq}{q}F'_{i}(q)
\frac{p}{q}D^{M}_{i}(\frac{p}{q}),
\end{equation}
where
\begin{equation}
F'_{i}(q)=\frac{1}{\beta L}\int^{qe^{\beta L}}_{q}dkkf_{i}(k),
\end{equation}
and $D^{M}_{i}$ is the FF of quark $i$ splitting into meson $M$.

In the following sections, we show the results of heavy flavored meson transverse momentum spectra.

\section{Productions of charmed mesons}
Anomalous $J/\psi$ suppression is regarded as an important tool to probe the hot dense matter or quark-gluon plasma (QGP) created in heavy ion collisions at RHIC energies \cite{QGP}. The study of open charmed mesons ($D$ and $\overline{D}$) may help to explain the suppression of charmonium states ($J/\psi$, $\chi_c$, $\psi'$). In Ref.\cite{pengru1}, we have obtained the fugacity of $c$ quark $\gamma_c=0.26$ and $v_T=0.3c$ by fitting the measured $J/\psi$ transverse momentum spectrum in the region of low $p_T$. Recently STAR collaboration have measured $D^0$ production at low transverse momentum ($p_T<2$ GeV$/c$) in Au+Au collisions at RHIC energy \cite{STAR}. To fit the experimental data with the thermal-thermal term of Eq.(\ref{TT}), the factor of $v_T$ is adjusted to be $v_T=0.42c$, which is a little larger than the fitted results in Ref.\cite{pengru1}, where $v_T=0.3c$. Since $J/\psi$ is produced at the early stage in the heavy ion collisions, and $D^0$ is produced later, it's reasonable that the radial flow velocity $v_T$ of $D^0$ is lager than that for $J/\psi$.

In Ref.\cite{pengru1}, the SPDs initiated by hard partons of charm quark and gluon have been achieved at fixed starting scale $Q=2m_c=3$ GeV$/c$. Other SPDs $S^{j}_{i}$ with $i=q, \overline{q}, g$ and $j=q, s, \overline{q}, \overline{s}$ ($q=u, d, s$) denoted by $K_{NS}$, $L$, $G$, $L_s$ and $G_s$ in Ref.\cite{SPD} are obtained at $Q=10$ GeV$/c$. In order to set the SPDs above at the same scale, we investigate the SPDs $K_{Ns}$ etc. at $Q=3$ GeV$/c$ with the method provided in Ref.\cite{TZG}. In calculating the thermal-shower ($\mathcal{TS}$) term with Eq.(\ref{TS2}), there are two kinds of shower partons. The sources of shower parton distributions of $c$ and $\overline{u}$ given by Eq.(\ref{shower}) are different. For charm shower partons the dominated contributions are from $i=c, g$ \cite{pengru1}, while for $\overline{u}$ shower partons they can come from $i=q,\overline{q}, g$. Other contributions to the involved shower partons are neglectable because of either lower contents in the shower or less number of initial hard partons. So the summation formula Eq.(\ref{shower}) of $\overline{u}$ shower parton distribution in Au+Au collisions can be expressed as
\begin{equation}
\mathcal{S}_{\overline{u}}(p)=\int\frac{dq}{q}[(K_{NS}(\frac{p}{q})+L(\frac{p}{q}))
F_{\overline{u}}(q)+L(\frac{p}{q})(F_{u}(q)+F_{d}(q)+F_{\overline{d}}(q))
+G(\frac{p}{q})F_{g}(q)].
\end{equation}
For $D^0(c\overline{u})$ meson, the RF is determined by Eq.(\ref{MRF}) with $a/b=1/5(a=1)$ as mentioned in Section 2.

\begin{figure}[tbph]
\includegraphics[width=0.4\textwidth]{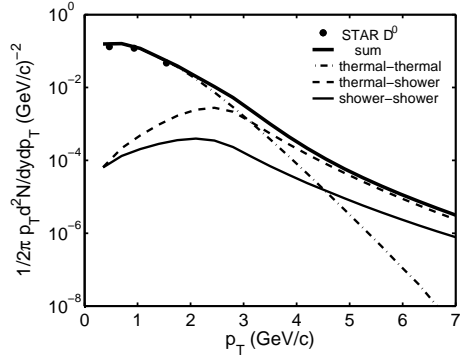}
    \caption{$D^0$ transverse momentum spectrum in central Au+Au collisions with the data at low transverse momentum \cite{STAR}.}
  \label{figD0}
\end{figure}

In Eq.(\ref{SS}) $i=g, c$ because the $D^0$ FFs of $u, d, s$ quarks are assumed to be zero at the starting scale \cite{BAKGK}. With the known FFs of $D^{D^{0}}_{g}$ \cite{pengru1} and $D^{D^{0}}_{c}$ \cite{BAKGK}, we can calculate the shower-shower ($\mathcal{SS}$) term. The results of $D^0$ production are shown in Fig.\ref{figD0}. In order to reflect the impact
of the momentum on the scale, the results of $\mathcal{TS}$ and $\mathcal{SS}$ terms are
multiplied by a factor of $1-e^{-p/2}$ which suppresses the low $p$ contributions.

\begin{figure}[tbph]
\includegraphics[width=0.4\textwidth]{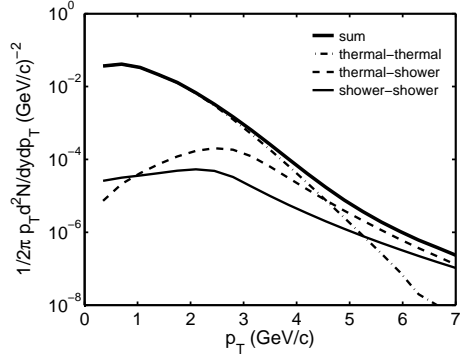}
    \caption{Predicted $D_s$ transverse momentum spectrum in central Au+Au collisions.}
  \label{figDs}
\end{figure}

We can predict $D_s$ production if we substitute $\overline{s}$ for $\overline{u}$ in the $D^0$ calculations. And the RF of $D_s$ is determined by Eq.(\ref{MRF}) with $a/b=m_s/m_c\simeq3/10$ ($a=3$). The SPD of $S^{s}_{c}(z)$ can be obtained with the FF of $c$ quark splitting into $K$ $D_{c}^{K}$ \cite{AKK} (review the method in detail in Refs.\cite{pengru1} and \cite{SPD}). And then we can calculate the FFs of $D^{D_s}_{c}$ and $D^{D_s}_g$ which are necessary in the calculation of $\mathcal{SS}$ term. The predicted results of $D_s$ production for 0-20\% central collisions are shown in Fig.\ref{figDs}. The contributions of $\mathcal{TS}$ terms are higher than those of $\mathcal{SS}$ terms at $p_T>0.4$ GeV$/c$ for $D^0$ or $p_T>1$ GeV$/c$ for $D_s$, because it is hard to produce $c$ shower partons in the jets created in the heavy ion collisions due to the large mass of $c$ quark.

\section{Productions of $b$-flavored mesons}
\begin{figure}[tbph]
\includegraphics[width=0.4\textwidth]{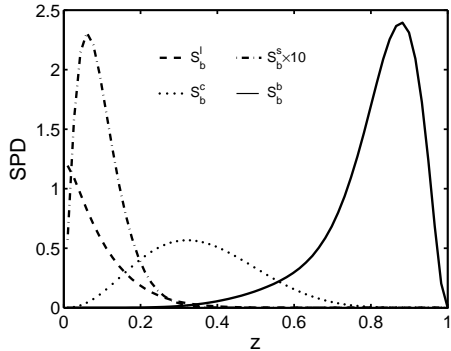}
    \caption{Shower parton distributions.}
  \label{figSPD2}
\end{figure}

The most remarkable difference between the situation of $b$-flavored mesons and charmed mesons is that all the FFs of $b$ quark used in this paper are obtained at starting scale of $Q=2m_b=10$ GeV$/c$. So we should try to get the SPDs initiated by $b$ quark with the method shown in Refs.\cite{pengru1} and \cite{SPD}:
\begin{equation}
\label{FF}
xD(x)=\int^x_0\frac{dx_1}{x_1}\int^x_0\frac{dx_2}{x_2}
\{S^{q}_{i}(x_{1}),S^{\overline{q'}}_{i}(x_{2})\}R_{M}(x_{1},x_{2},x),
\end{equation}
where
\begin{equation}
\{S^{q}_{i}(x_{1}),S^{\overline{q'}}_{i}(x_{2})\}\equiv\frac{1}{2}[S^{q}_{i}(x_{1})
S^{\overline{q'}}_{i}(\frac{x_{2}}{1-x_{1}})+S^{q}_{i}(\frac{x_{1}}{1-x_{2}})S^{\overline{q'}}_{i}(x_{2})]
\end{equation}
reflecting the symmetrization of the leading parton momentum fraction. The RFs $R_M(x_1,x_2,x)$ determined by Eq.(\ref{MRF}) for $b$-flavored mesons are denoted with $a/b=m_d/m_{\overline{b}}\simeq3/50$ ($a=3$) for $B$, $a/b=m_s/m_{\overline{d}}\simeq4.5/50$ ($a=4.5$) for $B_s$ and $a/b=m_c/m_{\overline{b}}\simeq3/10$ ($a=3$) for $B_c$, respectively. With the FFs $D(x)$ of $b$ splitting into $\pi$, $K$ \cite{AKK}, $D^0$ \cite{BAKGK} and $B$ \cite{bB}, we can get the results of $S^{l}_{b}$ ($l=u(\overline{u}), d(\overline{d})$), $S^{s}_{b}$, $S^{c}_{b}$ and $S^{b}_{b}$. When we talk about FF of $b\rightarrow B$, we have in mind that there are four fragmentation processes $\overline{b}\rightarrow B^{+}$, $\overline{b}\rightarrow B^{0}$, $b\rightarrow B^{-}$ and $b\rightarrow \overline{B}^{0}$. Strong assumption are made in Ref.\cite{bB} that the FFs of these four processes all coincide if the influence of the electroweak interactions is neglected. The results of these SPDs are shown in Fig.\ref{figSPD2}.

\begin{figure}[tbph]
\includegraphics[width=0.4\textwidth]{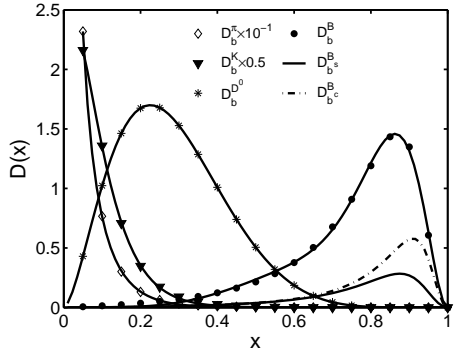}
    \caption{Fragmentation functions of $b$ splitting into $\pi$, $K$ \cite{AKK},$D^0$ \cite{BAKGK} and $B$ \cite{bB}. The predictions of $B_s$ and $B_c$ are shown with the curves without any symbols.}
  \label{figFFb}
\end{figure}

Again, applying the SPDs $S_b^b$ and $S_b^s$ ($S_b^c$) to Eq.(\ref{FF}) we can calculate the unknown FFs of $b$ splitting into $b$-flavored meson $B_s$ ($B_c$) further. We use the Peterson fragmentation function \cite{Peterson}
\begin{equation}
D_b(x)=N\frac{x(1-x)^2}{[(1-x)^2+\epsilon x]^2},
\end{equation}
which has been frequently applied in connection with the fragmentation of heavy quarks into their mesons. The two parameters are $N=0.0201$, $\epsilon=0.0177$ for $B_s$, and $N=0.0227$, $\epsilon=0.0098$ for $B_c$, respectively. The reproduced FFs with the parametrized SPDs and the predicted FFs are shown in Fig.\ref{figFFb}. As the results have shown, all the fits are excellent.

It's necessary to point out that $b$ shower parton initiated by gluon is not included in this paper, since the FFs of gluon to $b$-flavored mesons at $Q=10$ GeV$/c$ are not known. So we calculate the shower parton distribution of $b$ (Eq.(\ref{shower})) and the shower-shower ($\mathcal{SS}$) term (Eq.(\ref{SS})) with $i=b$. It's another difference in calculating $b$-flavored mesons from that for charmed mesons.

\begin{figure}[tbph]
\includegraphics[width=0.4\textwidth]{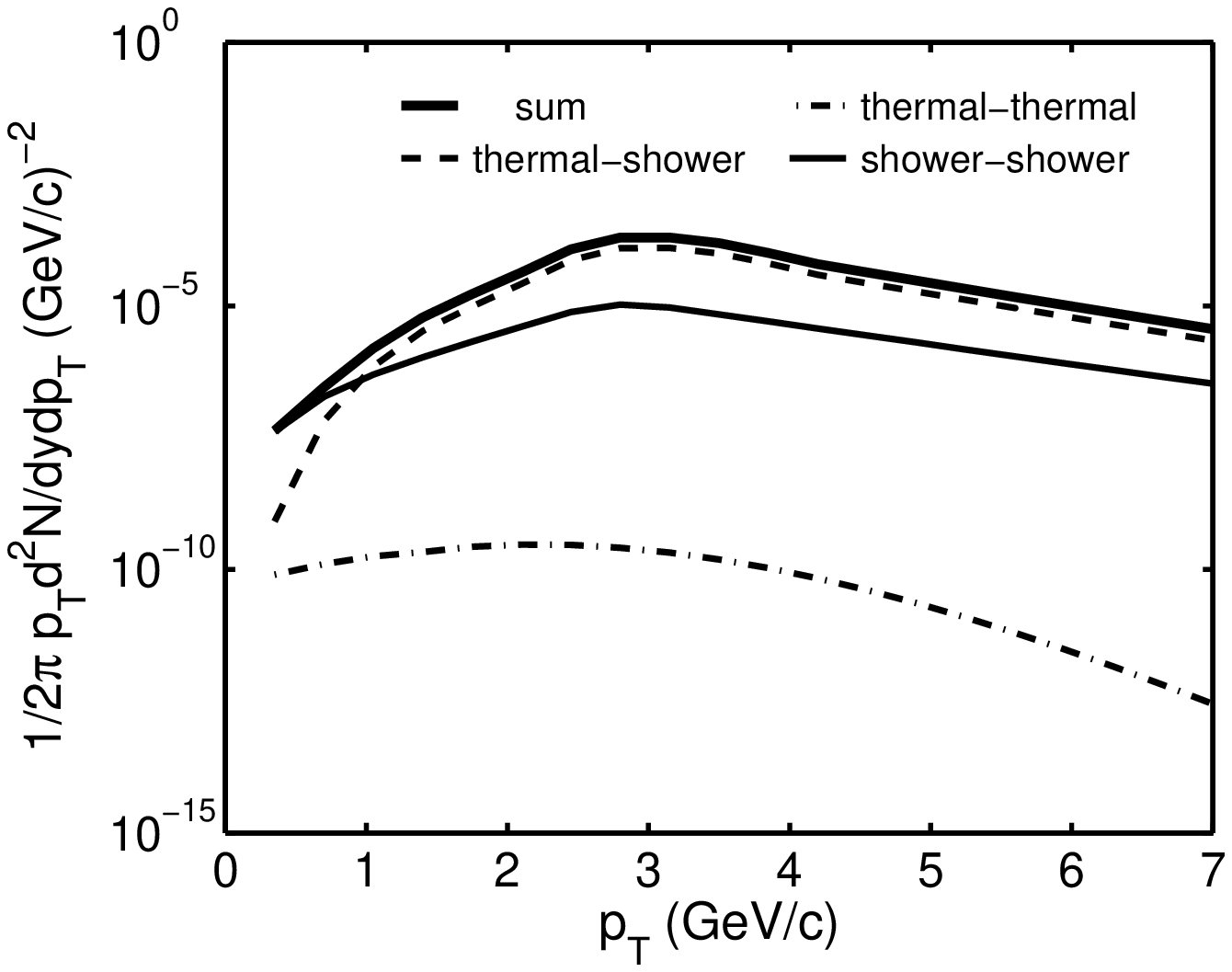}
    \caption{Predicted $B$ transverse momentum spectrum in central Au+Au collisions.}
  \label{figB}
\end{figure}

\begin{figure}[tbph]
\includegraphics[width=0.4\textwidth]{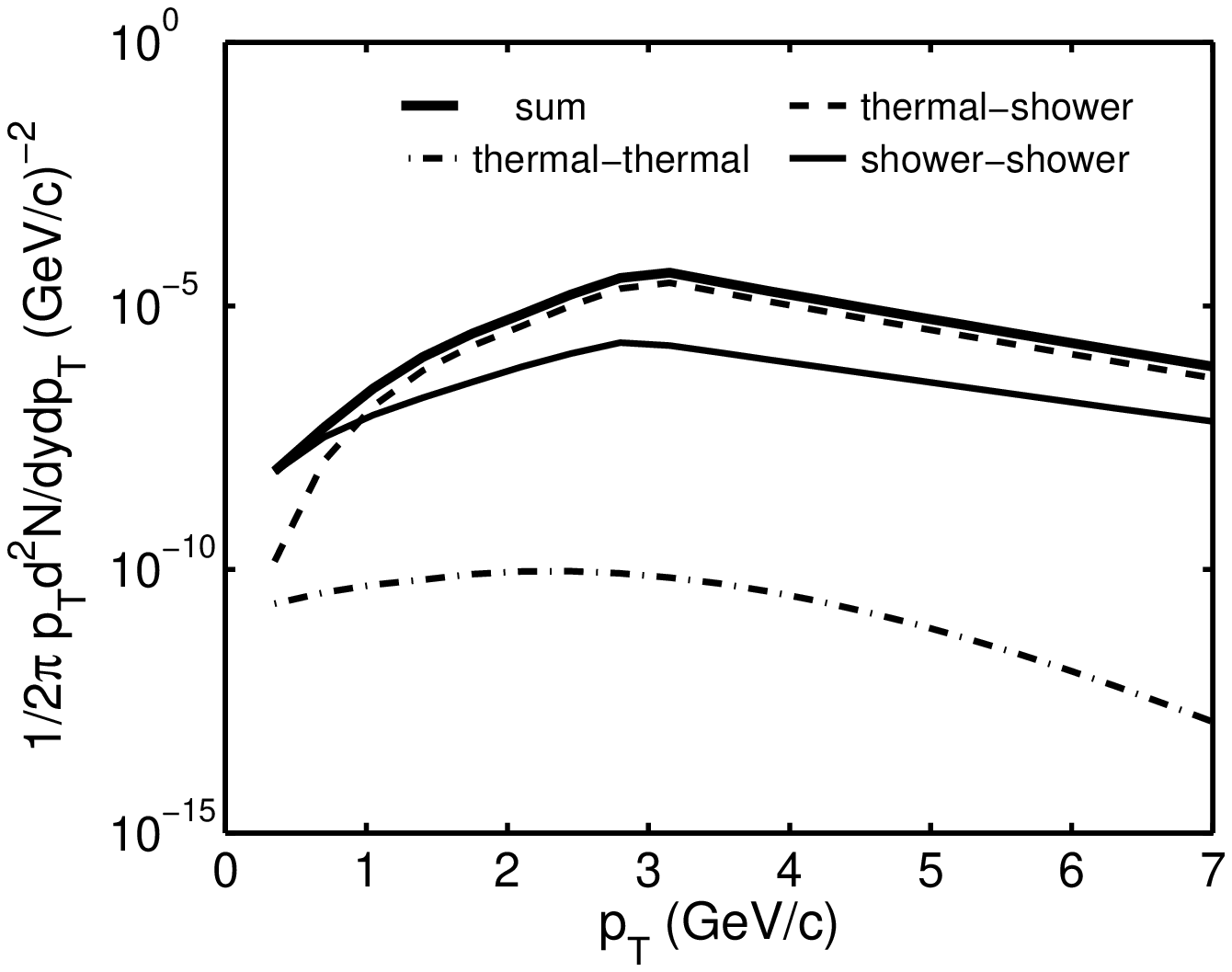}
    \caption{Predicted $B_s$ transverse momentum spectrum in central Au+Au collisions.}
\label{figBs}
\end{figure}

\begin{figure}[tbph]
\includegraphics[width=0.4\textwidth]{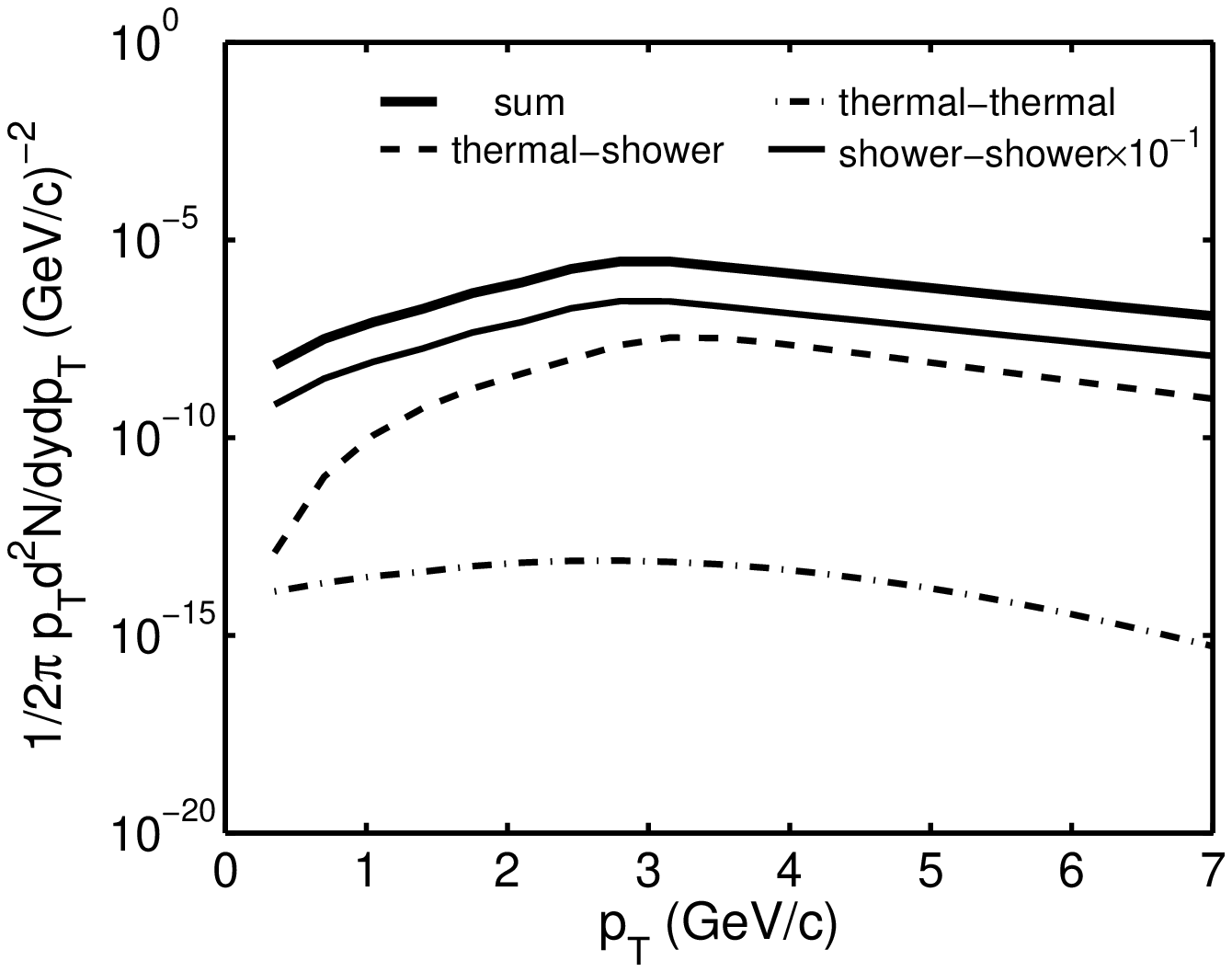}
    \caption{Predicted $B_c$ transverse momentum spectrum in central Au+Au collisions.}
  \label{figBc}
\end{figure}

\begin{figure}[tbph]
\includegraphics[width=0.4\textwidth]{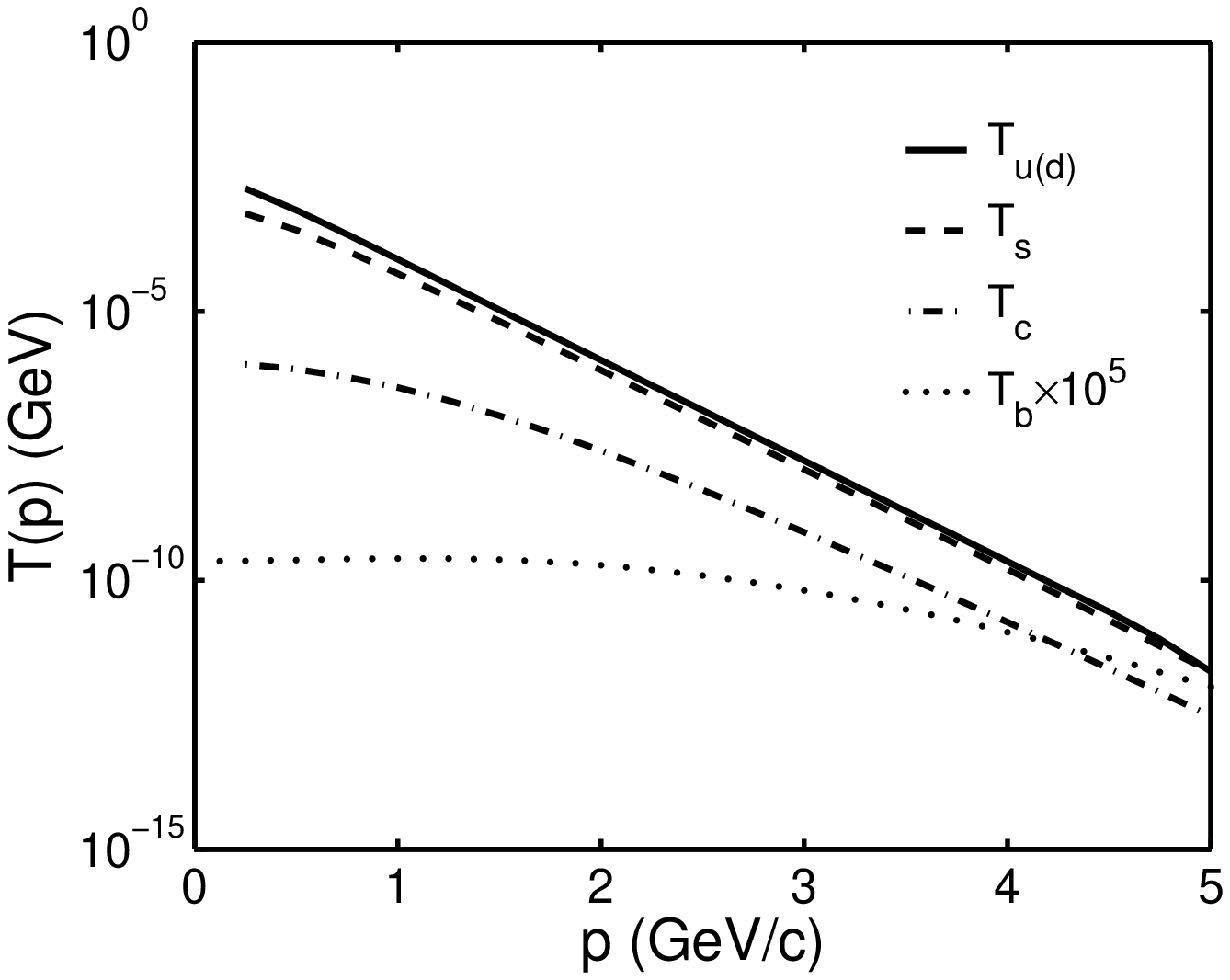}
    \caption{Thermal parton distributions.}
  \label{figTparton}
\end{figure}

We present the transverse momentum spectra of $b$-flavored mesons $B$, $B_s$ and $B_c$ for 0-20\% central collisions in Fig.\ref{figB}-Fig.\ref{figBc}. The results show that the contributions of $\mathcal{TT}$ are much lower than $\mathcal{TS}$ and $\mathcal{SS}$. The main reason is that the mass of $b$ quark $m_b=5$ GeV is much larger than those of light quarks and $c$ quark, which leads the much smaller thermal parton distribution. To see this point clearly, we exhibit the thermal parton distribution densities $\mathcal{T}(p)$ of $b$, $c$ with $\gamma_b=\gamma_c=0.26$ and other light quarks in Fig.\ref{figTparton} for the purpose of comparisons. The distribution of $b$ quark is almost eight orders magnitude lower than others. But this has not serious impact on the results of $\mathcal{TS}$ terms, because the contributions consist of two parts as shown in Eq.(\ref{TS2}). Take $B$ meson as an example, one part is $\mathcal{T}_{d}(p_1)\mathcal{S}_{\overline{b}}(p_2)$ and the other is $\mathcal{S}_{d}(p_1)\mathcal{T}_{\overline{b}}(p_2)$. It's obvious that the major component of the $\mathcal{TS}$ term comes from the first part, which is not affected by the lower thermal parton distribution of $b$. So the $\mathcal{TS}$ contributions of $B$ and $B_s$ productions are higher than $\mathcal{SS}$ in the region of $p_{T}>1$ GeV/$c$. While for $B_c$, the combined effect of the lower thermal parton distributions of $c$ and $b$ makes the $\mathcal{TS}$ contribution lower than $\mathcal{SS}$.

The fact that $\mathcal{TS}$ or $\mathcal{SS}$ contribution is much larger than $\mathcal{TT}$ contribution leads the different behavior of $p_T$ spectra for b-flavored mesons with that for charmed mesons. For $D^0$ and $D_s$ the dominant contributions come from $\mathcal{TT}$ terms at low $p_T$ because the thermal parton distribution of $c$ is higher than that of $b$. While for $b$-flavored mesons the primary behavior of the spectra are determined by $\mathcal{TS}$ or $\mathcal{SS}$ contribution.


\section{Meson production for different centralities}
\begin{figure}[tbph]
\includegraphics[width=0.4\textwidth]{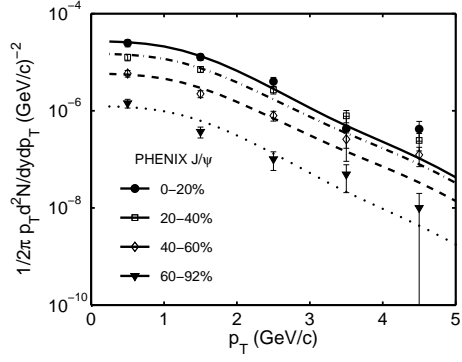}
    \caption{$J/\psi$ transverse momentum spectrum in Au+Au collisions for different centralities. The experimental data are from Ref.\cite{AA2}}
  \label{figJpsi1}
\end{figure}

In Ref.\cite{pengru1}, we have investigated the $J/\psi$ transverse momentum spectrum in central Au+Au collisions at midrapidity. In this paper we can predict the heavy flavored meson productions for non-central centralities. The fitted parameters involved can be obtained by fitting the experimental data of $J/\psi$ productions measured by PHENIX collaboration \cite{AA2}. The flow velocity $v_T$ and heavy quark fugacity $\gamma_c$ in Eq.(\ref{TT}) ($\mathcal{TT}$ term) are the parameters adjusted to fit the experimental data at low $p_T$ ($p_T<2$ GeV$/c$). We find that the data are fitted well with the changed $\gamma_c$ and the fixed $v_T=0.3c$ as can be seen in Fig.\ref{figJpsi1}. We suppose that fugacities for other quarks ($u$, $d$ and $s$) are proportional to those for $c$ quarks.

\begin{figure}[tbph]
\includegraphics[width=0.4\textwidth]{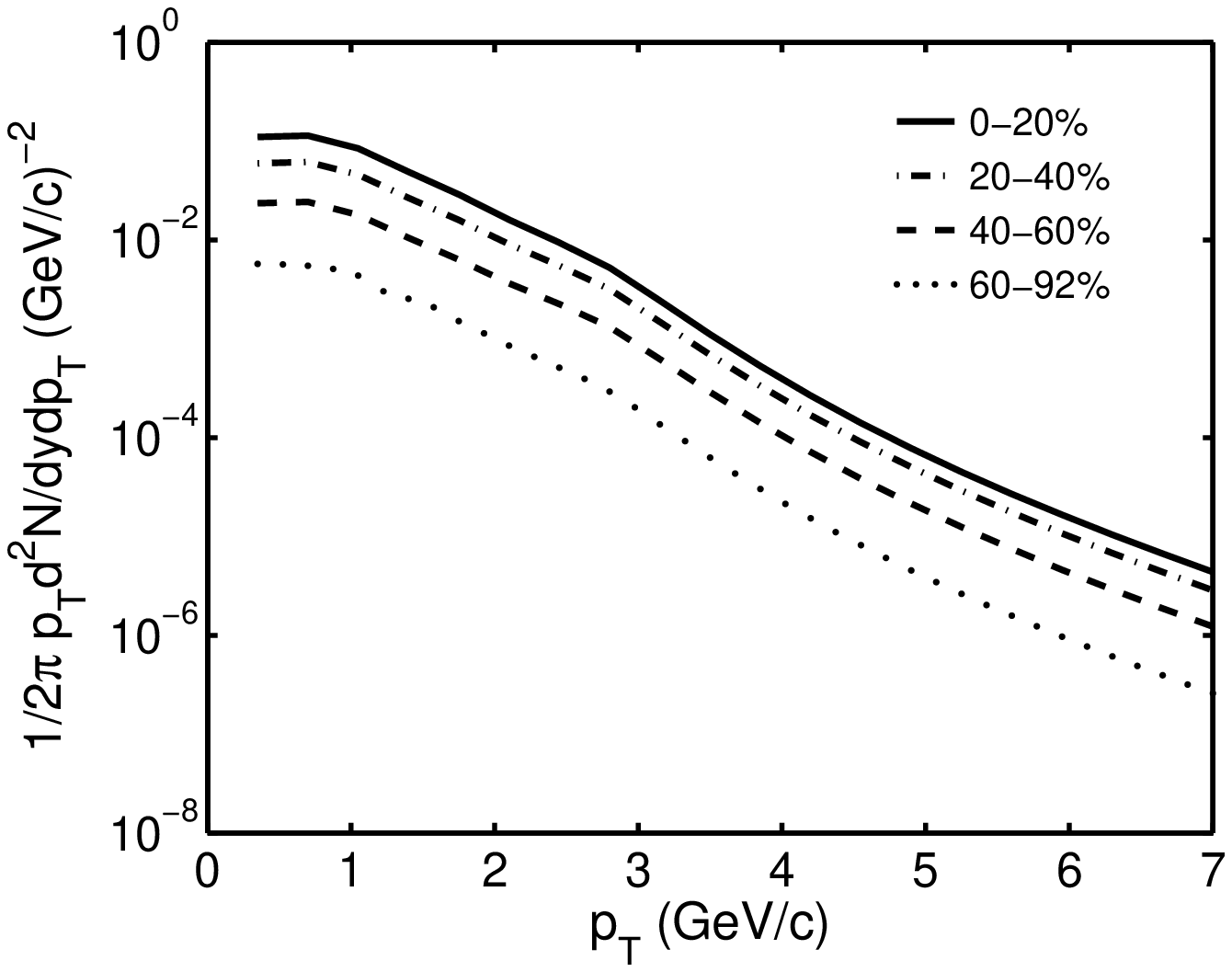}
    \caption{Predicted $D^0$ transverse momentum spectrum in Au+Au collisions for different centralities.}
  \label{figD01}
\end{figure}

\begin{figure}[tbph]
\includegraphics[width=0.4\textwidth]{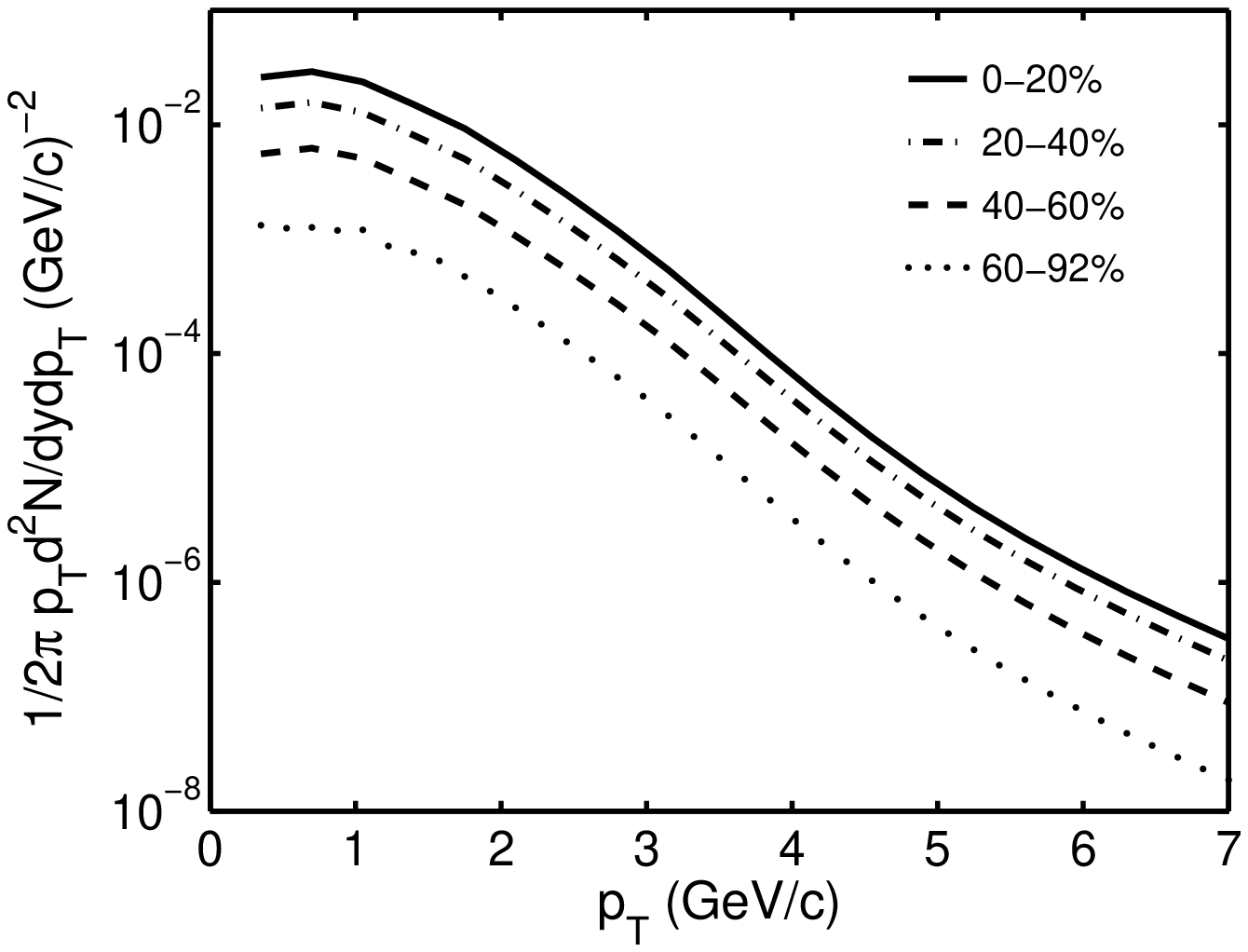}
    \caption{Predicted $D_s$ transverse momentum spectrum in Au+Au collisions for different centralities.}
  \label{figDs1}
\end{figure}

\begin{figure}[tbph]
\includegraphics[width=0.4\textwidth]{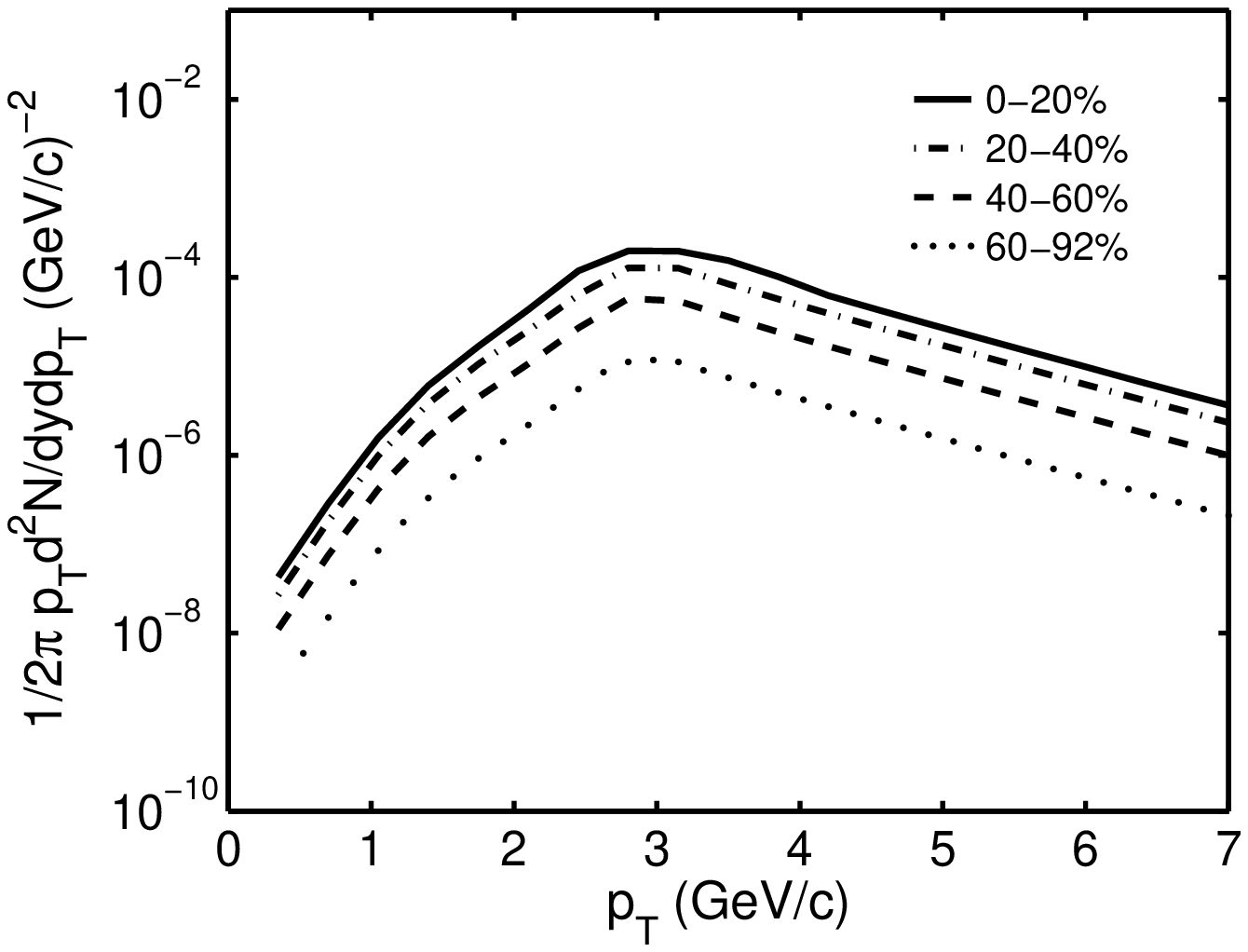}
    \caption{Predicted $B$ transverse momentum spectrum in Au+Au collisions for different centralities.}
  \label{figB1}
\end{figure}

\begin{figure}[tbph]
\includegraphics[width=0.4\textwidth]{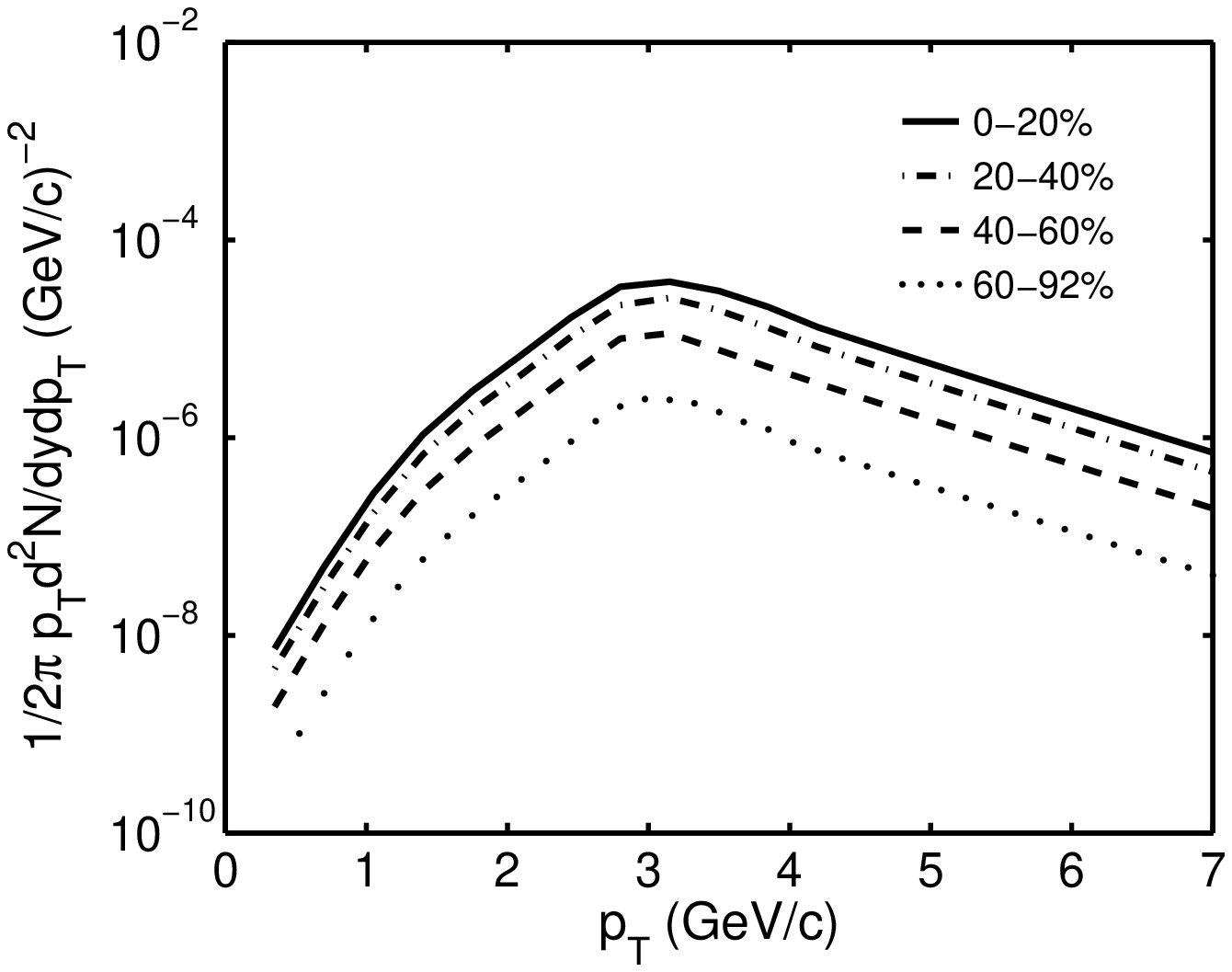}
    \caption{Predicted $B_s$ transverse momentum spectrum in Au+Au collisions for different centralities.}
  \label{figBs1}
\end{figure}

\begin{figure}[tbph]
\includegraphics[width=0.4\textwidth]{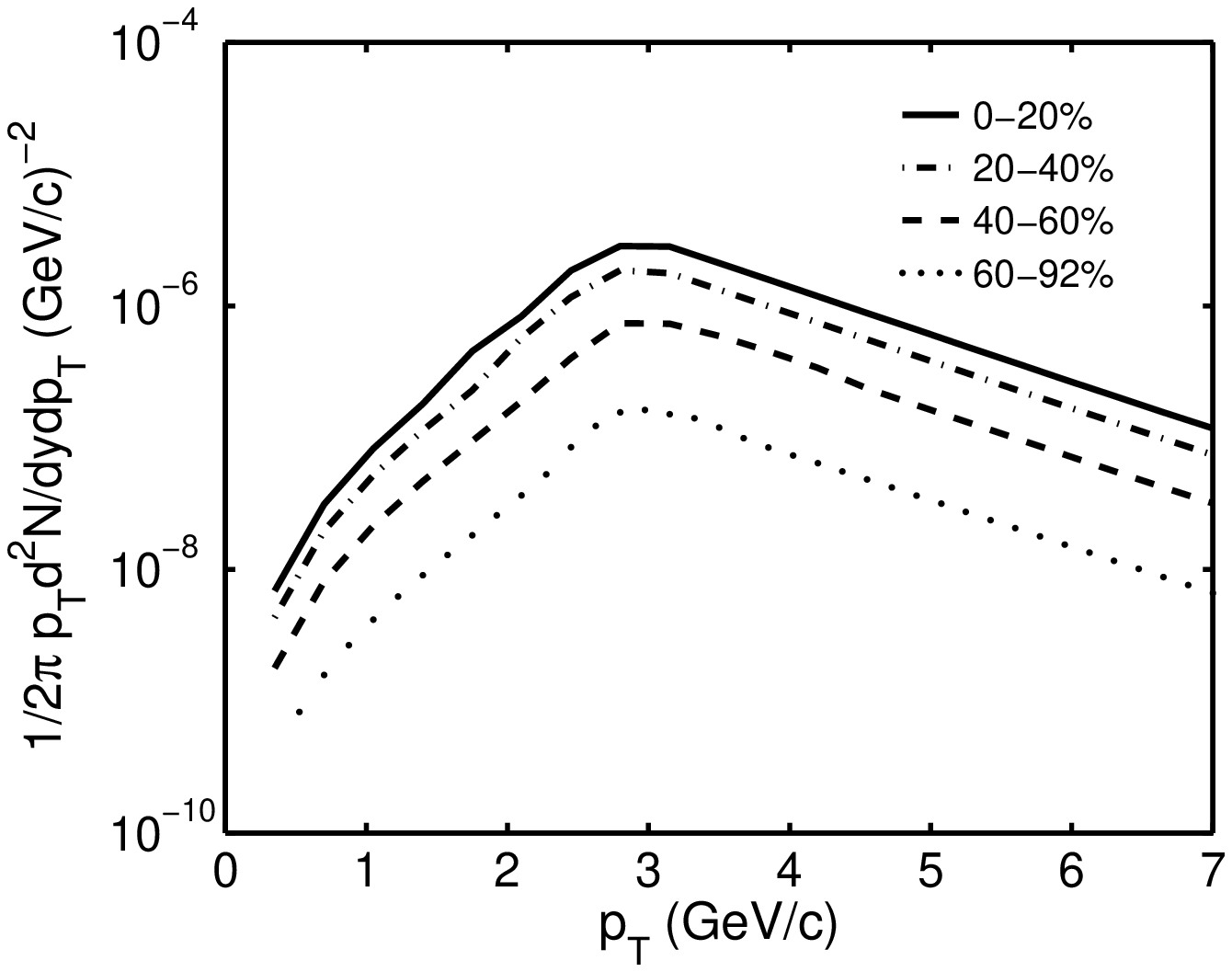}
    \caption{Predicted $B_c$ transverse momentum spectrum in Au+Au collisions for different centralities.}
  \label{figBc1}
\end{figure}

When we calculate the $\mathcal{TS}$ and $\mathcal{SS}$ terms, $\beta L$ standing for the energy loss factor in the heavy ion collisions should be regarded as a function of centrality. And it is determined by the inclusive distribution of $\pi^0$ for all centralities in Au+Au collisions \cite{energyloss}. The different values of $\gamma$ and $\beta L$ for different centralities are given in Table 1. For the hard parton distributions $f_i(k)$ is scaled by the number of binary collisions $N_{coll}$. With the changed values of $\gamma$, $\beta L$ and $f_i(k)$ for the corresponding centralities we predict the spectra of heavy flavored mesons as shown in Fig.\ref{figD01}-Fig.\ref{figBc1} for $D^0$, $D_s$, $B$, $B_s$ and $B_c$ for centrality bins 0-20\%, 20-40\%, 40-60\% and 60-92\%.

\begin{table}[h]
\caption {parameters for quark fugacities $\gamma$ and $\beta L$ for different centralities}
\begin{center}
\begin{tabular}{ccccc} \hline
 & $0-20\%$ & $20-40\%$ & $40-60\%$ & $60-92\%$  \\
\hline
$\gamma_c=\gamma_b$ & 0.26 & 0.19 & 0.12 & 0.06 \\
$\gamma_u=\gamma_d$ & 1.00 & 0.73 & 0.46 & 0.23 \\
$\gamma_s$ & 0.80 & 0.58 & 0.37 & 0.18 \\
$\beta L$ & 2.39 & 1.44 & 1.03 & 0.78  \\
\hline
\end{tabular}
\end{center}
\end{table}

\section{Conclusions}
We complete the SPDs initiated by the hard partons of heavy quarks $c$ and $b$ by making use of the FFs in the framework of the recombination model. With the parametrized SPDs, we reproduce the FFs of $c$ or $b$ splitting into charmed mesons or $b$-flavored mesons and predict other unknown FFs. The meson spectrum in relativistic heavy ion collisions is the summation of the recombination of thermal-thermal, thermal-shower and shower-shower partons, where the shower-shower recombination equals to the subprocess of hard scattering followed by parton fragmentation. As the spectra of the heavy flavored mesons have shown, the contributions from thermal-shower recombination are higher than those from shower-shower recombination in the region of $p_T>1$ GeV$/c$ except for the results of $B_c$. So the recombination of thermal-shower partons is important, of course, at every high $p_T$, shower-shower recombination will be the dominant process for the heavy flavored meson production.

\section*{Acknowledgements}
This work was supported in part by the National Natural Science Foundation of China under Grant Nos. 10635020 and 10775057, by the Ministry of Education of China under Grant No. 306022 and project IRT0624, and by the Programme of Introducing Talents of Discipline to Universities under No. B08033.

\end{document}